\documentclass[aps,prd,preprint,superscriptaddress,showpacs,floatfix,nobibnotes,nofootinbib]{revtex4-1}
\usepackage[usenames, dvipsnames]{color}

\usepackage{float}
\usepackage{latexsym}
\usepackage{amsmath}
\usepackage{amssymb}
\usepackage{graphicx}
\usepackage{hyperref}
\usepackage{setspace}
\usepackage{longtable}
\usepackage{enumitem}
\usepackage[bottom]{footmisc}

\usepackage[normalem]{ulem}
\usepackage[usenames, dvipsnames]{color}

\usepackage{setspace}
\usepackage{comment}

\usepackage{bm}
\usepackage{epsfig}
\newcommand{\bea}{\begin{eqnarray}}
\newcommand{\eea}{\end{eqnarray}}

\newcommand{\nn}{\nonumber}

\newcommand{\PP}{{\mathcal P}}

\begin{document}
\setlength{\baselineskip}{20pt}

\title{Evidence for universal flow and characteristics of early time thermalization in a scalar field model for heavy ion collisions
}

\author{Margaret E. Carrington}
\affiliation{Department of Physics, Brandon University,
Brandon, Manitoba R7A 6A9, Canada}
\affiliation{Winnipeg Institute for Theoretical Physics, Winnipeg, Manitoba, Canada}

\author{Wade N. Cowie}
\affiliation{Department of Physics, University of Manitoba, Winnipeg, Manitoba, R3T 2N2, Canada}
\affiliation{Winnipeg Institute for Theoretical Physics, Winnipeg, Manitoba, Canada}

\author{Gabor Kunstatter}
\affiliation{Department of Physics, University of Winnipeg, Winnipeg, Manitoba, R3M 2E9 Canada}
\affiliation{Department of Physics, Simon Fraser University, Burnaby, British Columbia, V5A 1S6 Canada}
\affiliation{Winnipeg Institute for Theoretical Physics, Winnipeg, Manitoba, Canada}

\author{Christopher D. Phillips}
\affiliation{Department of Physics, Brandon University,
Brandon, Manitoba R7A 6A9, Canada}

\today

\begin{abstract}

We study numerically the evolution of an expanding strongly self-coupled real scalar field. We use a conformally invariant action that gives a traceless energy-momentum tensor and is better suited to model the early time behaviour of a system such as QCD, whose action is also conformally invariant.
We consider asymmetric initial conditions and observe that when the system is initialized with non-zero spatial eccentricity, the eccentricity decreases and the elliptic flow coefficient increases. 
We look at a measure of transverse pressure asymmetry that has been shown to behave similarly to the elliptic flow coefficient in hydrodynamic systems and show that in our system their behaviour is strikingly similar. 
We show that the derivative of the transverse velocity is proportional to the gradient of the energy in Milne coordinates and argue that this result means that transverse velocity initially develops in the same way that it does in hydrodynamic systems. 

We conclude that some aspects of the early onset of hydrodynamic behaviour that has been observed in quark-gluon plasmas are seen in our numerical simulation of strongly coupled scalar fields. 
\end{abstract}

\maketitle


\section{Introduction}
\label{intro-sec}

The quark gluon plasma (QGP) has been observed experimentally to be well described by hydrodynamics at very early times ($1$ fm/c $\sim 3\times 10^{-24}$ seconds) after formation. This implies that some degree of thermalization must happen on even shorter time scales. Such rapid thermalization does not seem possible within a standard kinematical description of the QGP. A field theoretic approach therefore appears to be required, something that is extremely difficult in full Quantum Chromodynamics (QCD).

Progress has recently been made in \cite{Dusling:2010rm,Epelbaum:2011pc,Dusling:2012ig} (DEGV) in the context of a simple model that has some properties in common with QCD but is nonetheless considerably more tractable.  DEGV considered a real, self-interacting scalar field in four spacetime dimensions. Like QCD this theory is scale invariant and exhibits secular divergences in its perturbative solutions. These secular divergences are due to instabilities in the classical solutions.
DEGV used what is called the classical statistical approximation to resum the fastest growing divergences in this theory at each order in perturbation theory and showed that the scalar field system isotropizes when this resummation is performed \cite{Dusling:2012ig}. The only observables considered  were the energy and longitudinal/transverse pressures of the system, and it was found that the difference between the resummed longitudinal and transverse pressures tends towards zero.  
In a previous paper \cite{Carrington:2022ccm} we used the the classical statistical approximation to study angular momentum in a system of expanding scalar fields.  Some issues with the classical statistical approximation are discussed in \cite{Berges:2013lsa,Epelbaum:2014yja}. A completely different technique to resum the dominant contributions based on a 2-particle irreducible effective action approach can be found in \cite{Gelis:2024iar}.

In this paper we study the extent to which an expanding system of scalar fields behaves hydrodynamically, and how hydrodynamic behaviour develops with time. We start with a somewhat different theory than DEGV, one that we claim more closely models gluon behaviour without losing the inherent simplicity of a real scalar field. Specifically, we consider a real scalar that is non-minimally coupled to the background geometry so as to produce a conformally invariant action, i.e. one that is invariant up to a total derivative under local scale transformations as well as global ones. This action produces an  classical energy momentum tensor (EMT) that is manifestly symmetric, conserved and traceless on-shell without altering the equations of motion on a Minkowski background. The theory therefore shares with QCD its conformal symmetry and the resulting tracelessness of the EMT. In the rest of this paper we refer to the EMT obtained from the non-minimally coupled conformally invariant action as the conformal EMT. The conformal and canonical EMTs are almost indistinguishable at late times\footnote{Time is measured in units of the transverse coordinate lattice spacing.} but at early times some components of the canonical EMT are characterized by rapid fluctuations that are absent in the conformal EMT.

We use the conformal EMT to study some observables that are related to the transverse asymmetry of the system of fields. 
Transverse asymmetry has been studied for many years in the context of quark-gluon plasmas. In a relativistic non-central collision there is an initial spatial asymmetry that rapidly decreases with time, which means anisotropic transverse momentum can develop only at very early times. Since anisotropic flow is sensitive to the system's properties very early in its evolution, it can provide direct information about the early stages of the system. 
We find that if our system of scalar fields is initialized with spatial asymmetry, the asymmetry is transmitted into the momentum field. This behaviour mimics what is seen in heavy ion collisions and is commonly considered an indication of the onset of some kind of hydrodynamic behaviour. 
We also look at a measure of transverse pressure asymmetry that has been shown to behave similarly to the elliptic flow coefficient in quark gluon plasmas. We show that in our system the behaviour of these two quantities is qualitatively similar. 
Finally we investigate a  proposal \cite{Vredevoogd:2008id} that under very general conditions the derivative of the transverse velocity will be proportional to the gradient of the energy, and discuss the significance of the result.  

This paper is organized as follows.
In section \ref{formalism-sec} we discuss the classical theory and the definition of the energy-momentum tensor (EMT). In section \ref{resum-sec} we describe the resummation method and give some details of the numerical procedure. We explain how the discretization of the equations is done, and our choice of boundary conditions and initial conditions. 
In section \ref{observ-sec} we discuss how the energy density and the transverse and longitudinal pressures are obtained from the EMT. We derive some equations that characterize the hydrodynamic behaviour of the system, and we define functions that measure the eccentricity,  elliptic flow and transverse pressure asymmetry of the system. 
In section \ref{results-sec} we present our results and section 
\ref{sec-conclusions} concludes with a summary and some observations. Some useful equations are collected in Appendix \ref{A-sec}.

Throughout this paper the spacetime is taken to be Riemannian with signature $(+,-,-,-)$. 
In addition to Minkowski coordinates $(t,z,x,y)$ we will also use Milne coordinates $(\tau,\eta,x,y)$, where $\tau$ is the Bjorken time and $\eta$ is spacetime rapidity. We choose units such that $c=k_B=\hbar=1$, where $c$ is the speed of light in vacuum, $k_B$ is the Boltzmann constant, and $\hbar$ is the Planck constant divided by $2\pi$.
\section{The classical theory}
\label{formalism-sec}
\subsection{Canonical EMT}

The Lagrangian density used by DEGV is that of a massless real scalar field\footnote{We write everything covariantly since this will be useful when we switch to Milne coordinates, which are related to Minkowski coordinates by a non-linear coordinate transformation}:
\bea
{\cal L} = \frac{1}{2}\nabla^\mu \varphi \nabla_\mu \varphi - V \text{~~with~~}
V = \frac{g^2}{4!}\varphi^4\,.
\label{L-min}
\eea
The equation of motion is 
\bea
\nabla_\mu \nabla^\mu\varphi + \frac{g^2}{3!} \varphi^3 = 0 ~~
\to ~~ \varphi \Box \varphi + 4V = 0\,.
\label{eom-min}
\eea
The canonical EMT is
\bea
T_{\rm can}^{\mu\nu} = \frac{\delta{\cal L}}{\delta(\nabla_\mu \varphi)}\nabla^\nu \varphi - g^{\mu\nu}{\cal L}
 = \nabla^\mu \phi \nabla^\nu\phi - g^{\mu\nu}\left[\frac{1}{2}\nabla^\alpha\phi \nabla_\alpha\phi - \frac{g^2}{4!}\phi^4\right]\,. \label{Tmunu-can}
\eea
Its divergence is 
\bea
\nabla_\mu T_{\rm can}^{\mu\nu} 
 &=& (\Box\varphi) (\nabla^\nu \varphi) + (\nabla_\mu)(\nabla^\mu\nabla^\nu\varphi) - g^{\mu\nu} (\nabla_\alpha\varphi)(\nabla_\mu\nabla^\alpha\varphi) + g^{\mu\nu}\frac{g^2}{3!}\varphi^3 \nabla_\mu\varphi \nn \\
&=& (\Box\varphi) (\nabla^\nu \varphi)+ \frac{g^2}{3!}\varphi^3 \nabla^\nu\varphi =0
\label{div}
\eea
where we used the equation of motion in the last line. 
The conservation law (\ref{div}) implies  the conservation on shell of the momentum four-vector defined by
\bea
{\PP}^\mu = \int d^3 x\sqrt{-g} T^{\mu0}
\label{eq:four-momentum}
\eea
provided the metric does not explicitly depend on the spacetime coordinates\footnote{This condition is not satisfied in Milne coordinates.} and the fields vanish on the boundaries.

The trace of the canonical EMT is
\bea
T^\mu_{{\rm can}~\mu} &=& -\nabla_\mu \varphi \nabla^\mu \varphi + 4 V \nn \\
 &=& -\nabla_\mu \varphi \nabla^\mu \varphi  -\varphi \Box\varphi = -\nabla_\mu (\varphi \nabla^\mu \varphi)
\label{trace-can}
\eea
using the equation of motion to go from the second line to the third.  
The canonical EMT is therefore not traceless. 
The QCD canonical EMT is also not traceless, and additionally it is not symmetric or gauge invariant.  These problems can be remedied using several different methods (see \cite{Blaschke:2016ohs} for a detailed discussion). In this work we define the EMT from a conformally invariant scalar field action. It is manifestly traceless due to the conformal invariance of the action. It is also conserved and symmetric.

\subsection{Covariant EMT}
\label{conf-sec}

In this section we derive the conformal EMT for the real massless scalar field that, like the EMT used for QCD, is traceless.
 In section \ref{compare-sec} we explain further why the definition given in this section  is a better choice for our purposes.

The real scalar field Lagrangian (\ref{L-min}) is invariant under constant scale transformations
in which the spacetime metric and scalar field transform as:
\bea
g_{\mu\nu}\to\Omega g_{\mu\nu} \text{~~and~~}
\varphi \to \Omega^{-\tfrac{1}{2}}\varphi
\label{eq:CT}
\eea
with constant $\Omega$.
It is however not  invariant under local scale transformations (i.e. conformal transformations) for which $\Omega=\Omega(x)$ is an arbitrary function of the spacetime coordinates that goes to unity on the boundaries.
The Lagrangian density of QCD is conformally invariant and to incorporate as many of the fundamental properties of QCD into our theory as possible, we  start with a conformally invariant theory. The Lagrangian density is \cite{PT}:
\bea
{\cal L}_\xi = \frac{1}{2}\sqrt{-g}(g^{\mu\nu}\nabla_\mu \nabla_\nu \varphi - \xi R \varphi^2 -V)
\label{L-New}
\eea
where $R$ is the scalar curvature of the spacetime.  The choice $\xi=0$ gives the original Lagrangian (\ref{L-min}) and is called the minimally coupled theory. In four dimensions with $\xi=1/6$ the Lagrange density changes by a total derivative under the transformation (\ref{eq:CT}) with non-constant $\Omega(x)$.

The equation of motion for the scalar field is
\bea
\left(\Box +\frac{R}{6} + \frac{g^2}{3!}\varphi^2\right)\varphi = 0\,.
\label{eom-cf}
\eea
The metric is treated as a fixed background field and is therefore not dynamical. When evaluated on a Minkowski background (zero curvature), eq.~(\ref{eom-cf}) reduces to the equation of motion   (\ref{L-min}) of the  original theory. 

The requirement that the action be invariant under a general coordinate transformation $x^\mu \to x'^\mu - \epsilon^\mu(x)$ leads to the definition of a conserved, symmetric EMT\cite{PT}
\bea
T^{\mu\nu} = -\frac{2}{\sqrt{-g}}\frac{\delta {\cal L}}{\delta g_{\mu\nu}}\,.
\label{Tmunu-cf}
\eea
 One advantage of the definition in (\ref{Tmunu-cf}) is that the tensor $T^{\mu\nu}$ is always symmetric and preserves the symmetries of the Lagrangian density. This is not necessarily true for the definition in equation (\ref{Tmunu-can}). 
The resulting  EMT for the Lagrange density in (\ref{L-New}) is
\bea
T^{\mu\nu} &=& \nabla^\mu \phi \nabla^\nu\phi - g^{\mu\nu}\left[\frac{1}{2}\nabla^\alpha\phi \nabla_\alpha\phi - \frac{g^2}{4!}\phi^4\right]
+\frac{1}{6}\left[g^{\mu\nu} \Box  - \nabla^\mu\nabla^\nu  \right]\varphi^2 \nn \\
&-&\frac{1}{6}(R^{\mu\nu}-\frac{1}{2}g^{\mu\nu}R)\varphi^2\,. \label{Tmunu-hat}
\eea
It is straightforward to verify that $T^{\mu\nu}$ is traceless and divergenceless in four dimensions. In flat spacetime where $R=0$ we can write this result in the form
\bea
&& T^{\mu\nu} = T_{\rm can}^{\mu\nu} + T_{\rm ex}^{\mu\nu} \nn \\
&& T_{\rm ex}^{\mu\nu} = \frac{1}{6}\left[g^{\mu\nu} \Box  - \nabla^\mu\nabla^\nu  \right]\varphi^2
\label{Tmunu-ex}
\eea
where $T_{\rm can}^{\mu\nu}$ is the canonical EMT (\ref{Tmunu-can}).

\subsection{Milne coordinates}
\label{milne-sec}
Milne coordinates are defined as
\bea
    \tau = \sqrt{t^2-z^2} \text{~~~and~~~} \eta = \frac{1}{2}\ln\left(\frac{t+z}{t-z}\right)
\eea
which gives $t=\tau \cosh(\eta)$ and $z=\tau\sinh(\eta)$.
The metric is
 \bea
g = (1,-\tau^2,-1,-1)_{\rm diag}\,.
 \eea
The coordinate $\eta$ is called the rapidity and depends only on the slope $v_z=\tfrac{z}{t}$. 
A change of $\eta$ corresponds to a boost by velocity $v_z$. It is therefore also  referred to as a boost coordinate. 
 
Milne coordinates are particularly well suited to study a high energy collision of nuclei moving along the $z$-axis. Surfaces of constant $\eta$ represent  timelike surfaces moving away from the origin at constant velocity and a box with boundaries at fixed $\eta$ expands in the $z$-direction with time. In this sense Milne coordinates  mimic the kinematics of a high energy collision. In our notation a ``dot'' indicates a derivative with respect to $\tau$, and $\vec\nabla_{\perp}$ is the transverse gradient operator.
The Riemann tensor (and the Ricci scalar) are both zero and therefore the equation of motion (\ref{eom-cf}) is the same as the one obtained from the minimally coupled theory (\ref{eom-min}). In Milne coordinates it takes the form 
\bea
\ddot{\phi}(\tau,\eta,\vec x_\perp)+\frac{1}{\tau}\dot{\phi} - \frac{1}{\tau^2}\partial^2_{\eta}\phi - \nabla^2_{\perp}\phi +\frac{g^2}{6}\phi^3 = 0\,.
\label{eom-milne}
\eea
In Appendix \ref{A-sec} we write all components of the two EMTs  (\ref{Tmunu-can}) and  (\ref{Tmunu-hat}) in Milne coordinates.

\section{Resummation of quantum fluctuations}
\label{resum-sec}

Observables in the scalar theory exhibit  secular divergences at next-to-leading order if they are calculated in a loop expansion. These divergences originate from instabilities of the classical solutions. The problem can be cured using the classical statistical approximation, which amounts to averaging over a Gaussian ensemble of initial conditions \cite{Dusling:2010rm,Dusling:2012ig}. This procedure implements a resummation scheme that collects the leading secular terms at each order of an expansion in the coupling constant\footnote{
The resulting EMT has an ultraviolet divergence corresponding to a vacuum contribution but this can be removed by repeating the calculation with the background field set to zero and subtracting the results. This vacuum subtraction has been done for all the calculations presented in this paper.}. 
We briefly describe the structure of the calculation. Further details can be found in \cite{Dusling:2012ig,Carrington:2022ccm}. 

\subsection{Fluctuations}
\label{fluc-sec}

We write the initial field at some small but non-zero\footnote{The initial time $\tau_0$ must  be small in order to describe a system expanding from essentially zero volume ($z=0$ at $t=0$) but it cannot be exactly zero due to the coordinate singularity at $\tau=0$.  One can check that the value chosen for this small initial time does not change the results at finite times.} $\tau=\tau_0$
as the sum of a background field contribution,  $\varphi(\tau_0,\vec x_\perp)$, and a fluctuation, $\alpha^{(\gamma)}(\tau_0,\eta,\vec x_\perp)$
\bea
\phi^{(\gamma)}(\tau_0,\eta,\vec x_\perp) = \varphi(\tau_0,\vec x_\perp) + \alpha^{(\gamma)}(\tau_0,\eta,\vec x_\perp)\,.
\label{phi-0}
\eea
The field $\phi^{(\gamma)}(\tau,\eta,\vec x_\perp)$ at finite proper time is obtained by solving equation (\ref{eom-milne}) with the initial condition (\ref{phi-0}).
We then obtain resummed values of  observable quantities by averaging over the  ensemble:
\bea
\langle O(\tau,\eta, x_\perp) \rangle = \frac{1}{N_\gamma}\sum^{N_\gamma}_{\gamma=1} O[\phi^{(\gamma)}(\tau,\eta, x_\perp)] \,.\nn
\eea
We explain below how the two terms in (\ref{phi-0}) are calculated and the meaning of the index $\gamma$. 

The field $\varphi(\tau_0,\vec x_\perp)$ 
is constructed from solutions of the classical equation of motion (\ref{eom-milne}) at $\tau_0$. 
It is assumed independent of the spatial rapidity $\eta$. 
The motivation behind this assumption is the physical picture of a heavy ion collision in which  the nuclei pass through each other without significant slowing. 
The resulting velocity distribution has a property called Bjorken
boost-invariance - which is that the longitudinal velocity $v_z$ of frames locally comoving with
the fluid is related to their spacetime position by $v_z = z/t$. A fluid with this velocity
distribution will look the same in all longitudinally comoving fluid elements.
We  also assume that at very early times the dynamics of the system is dominated by expansion and  we  therefore drop the interaction term in the equation of motion for the purposes of determing $\varphi(\tau_0,\vec x_\perp)$. The resulting equation is linear in $\varphi$ and separating variables we have that at fixed $\tau_0$ the background field can be written as a sum over plane wave mode functions. The specific combinations of plane waves that we use is discussed in section \ref{ics-sec}. 

The fluctuation $\alpha$ that is added to the background field in equation (\ref{phi-0}) carries an index $\gamma$ that indicates a Gaussian ensemble of $N_\gamma$ initial fluctuations defined as
\bea
&& \alpha^{(\gamma)}(\tau_0,\eta,\vec x_\perp) = \int dK \big[c^{(\gamma)}_K a_K+c^{(\gamma)*}_K a_K^*\big]\,\label{phi-1}
\eea
where the index $K$ labels the momentum variables  $(\nu,\vec k_\perp)$ that are conjugate to the coordinate-space variables   $(\eta,\vec x_\perp)$, respectively. 
The notation $c^{(\gamma)}_K$ denotes an element in a Gaussian distributed ensemble of $N_\gamma$ random numbers, with variance
\bea
\langle c^*_K c_L\rangle &:=& \sum_{\gamma=1}^{N_\Gamma} c^{(\gamma)*}_K c^{(\gamma)}_L = \frac{1}{2}\delta_{KL}\,.
\label{variance}
\eea 
The momentum space integration measure is
$
dK = d\nu d \vec k_\perp/ (2\pi)^3
$
and the delta function in equation (\ref{variance}) is defined so that $\int dK \delta_{KL} = 1$.
The initial mode functions $a_K\equiv a_{\nu \vec k_\perp}(\tau_0,\eta,\vec x_\perp)$ are obtained from the linearized equations of motion
\bea
\ddot a_K + \frac{1}{\tau}\dot a_K -\frac{1}{\tau^2} \partial^2_{\eta} a_K -\Delta_\perp a_K +\frac{g^2}{2} \varphi^2(\tau_0,\vec x_\perp) a_K =0\,
\label{eqnx1}
\eea
and normalized so that 
$\int dK (a_K,a_L) = 1$ with 
\bea
(a_K,a_L) = i \tau \int d\eta \int d^2 \vec x_\perp \, \big(a_K^* \partial_\tau a_L - (\partial_\tau a^*_K)a_L \big)\,.
\label{inner}
\eea
Separating variables and performing the normalization one finds \footnote{The time dependent part of the equation is second order and has two independent solutions. We use only the one that has positive frequency behaviour at large times which is  called the second Hankel function.
}
\bea
a_K&\equiv& a_{\nu \vec k_\perp}(\tau_0,\eta,\vec x_\perp) = \frac{1}{2}\sqrt{\pi} e^{\pi\nu/2}\,e^{i\nu\eta}\chi_{\vec k_\perp}(\vec x_\perp)  H_{i\nu}(\lambda_{\vec k_\perp}\tau_0) \,\label{bkg}
\eea
where the functions  $\chi_{\vec{k}_\perp}(\vec x_\perp)$ are solutions of the eigenvalue equation
\bea
\left[-\Delta_\perp+\frac{g^2}{2} \varphi^2(\tau_0,\vec x_\perp)\right]\chi_{\vec k_\perp}(\vec x_\perp) = \lambda^2_{\vec k_\perp}\chi_{\vec k_\perp}(\vec x_\perp)\,.
\label{chi-1}
\eea
When $\tau\to 0$ the  Hankel function in (\ref{bkg}) oscillates like $e^{\pm i \tau \nu}$ and the derivative diverges. As explained at the beginning of this section, we start the evolution at a small positive time $\tau_0=10^{-2}$. 
To calculate the Hankel function at this initial time we use an asymptotic series to find it at the smallest time
for which the series converges to the desired accuracy (numerically we truncate the series when successive terms are less than $10^{-9}$), and 
then use adaptive fifth order Runge-Kutta to evolve it to the initial time $\tau_0$. 

\subsection{Discretization}

We discretize  in both directions in the transverse plane with $L$ grid points and lattice spacing set to one, which means we define all dimensionful quantities in terms of the transverse lattice grid spacing.  
The rapidity variable $\eta$ is discretized with $N$ grid points and lattice spacing $h$. 
We consider a unit slice of rapidity with $\eta\in(-1/2,1/2)$ and $h=1/N$.
The discretized version of equation (\ref{chi-1}) is 
\bea
&& D_{ij;kl}\;\chi_{kl} = \lambda^2 \chi_{ij} 
\label{chi-2}
\eea
with
\bea
&& D_{ij;kl} = (4+V^{\prime\prime}_{ij})\delta_{ik}\delta_{jl} - (\delta_{i+1~k}+\delta_{i-1~k})\delta_{jl} - \delta_{ik}(\delta_{j+1~l}+\delta_{j-1~l})\,.
\label{D-mat}
\eea 
Since $D$ is a rank 4 tensor with $L^4$ components, we obtain $L$ eigenvalues $\lambda^e$ with $e\in(1,L^2)$ and $L^2$ eigenfunctions $\chi_{ij}^e$ which are normalized $
\sum_{ij} \chi^{*e}_{ij}\chi^{\bar e}_{ij} = L^2\,\delta^{e\bar e}
$. 
To discretize the longitudinal variables we note that the constraint 
$\partial_\eta^2 e^{i\nu\eta} = -\nu^2 e^{i\nu\eta}$
gives
\bea
\varepsilon_v := \nu = \frac{2}{h}\sin\left(\frac{\pi v}{N}\right)\,
\eea
and we replace $\nu \to \varepsilon_v$ in every factor $e^{\pi\nu/2}$. For the complex exponential we use
$e^{i\nu\eta} \to e^{\frac{2\pi i v n}{N}}$ and the integral over $\nu$ becomes a sum over $v$ using $
\int d\nu/(2\pi)  \to 1/(Nh) \sum_{v}^{N}$. 

Combining these results we find the discretized versions of equations (\ref{phi-0}, \ref{variance}, \ref{bkg}): 
\bea
&& \alpha_{nij}(\tau) =  \frac{1}{N L^2 h} \sum_{v=1}^{N}  \sum_{p=1}^{L^2} \left[c_{vp} a^{vp}_{nij}(\tau)\, + \text{c.c.}\right]\, \nn \\
&& a^{vp}_{nij}(\tau) = \frac{1}{2}\sqrt{\pi} e^{\frac{2\pi i v n}{N}}\,\chi_{ij}^p \, e^{\pi\nu/2} H^{(2)}_{i\nu}(\lambda_{\vec k_\perp}\tau) \, \nn \\[2mm]
&& \langle c_{ve}c^*_{u\tilde e}\rangle = \frac{1}{2}N L^2 h \delta_{vu}\delta_{e\tilde e}\,.
\eea
To verify that the discretization is done correctly we have checked  the discretized version of the normalization condition (\ref{inner}). 

We use periodic boundary conditions which means that the indices $(i,j)$ that correspond to the transverse spatial coordinates are defined modulo $L$, and the index $n$ for the rapidity is modulo $N$. 
The boundary conditions satisfy the self-adjointness condition
\bea
&& \nabla_F \phi(x) \equiv \phi(i+1)-\phi(i) \nn \\
&& \nabla_B \phi(x) \equiv \phi(i)-\phi(i-1) \nn \\
&& \sum_i f(i)\big(\nabla_F g(i)\big) = - \sum_i \big(\nabla_B f(i)\big) g(i) \,. \nn
\eea
We use forward derivatives, defined as $\partial_x f(x) \to f(i+1)-f(i)$, so that the integral of a total derivative term like the last term in equation (\ref{trace-can}) will be zero. 

\subsection{Initialization}
\label{ics-sec}

As explained in section \ref{fluc-sec}, the initial background field $\varphi(\tau_0,\vec x_\perp)$ can be written as a sum of plane wave mode functions of the form $\cos(\vec k_\perp \cdot \vec x_\perp)$. In this section we discuss how to choose these functions and construct the sum. 
Ideally we would like to consider initializations that correspond to collisions of sources with specified radius colliding with a specified impact parameter, and we could do this by constructing a wave-packet of  transverse plane waves that corresponds to a collision with some specific geometry \footnote{Since the sources that describe the colliding projectiles have support only on the light cone (\ref{eom-milne}), they do not directly drive the field evolution, but instead provide information on how to construct this wave-packet. 
In a Colour Glass Condensate (CGC) description of a heavy ion collision, the background field would contain a distribution of momentum modes up to the saturation momentum.}. 
This calculation is considerably more difficult than what has been done so far with the method of DEGV. 
For the purposes of studying the isotropization of the longitudinal and transverse pressures, it is sufficient to consider a classical background field that consists of only one mode, which corresponds to sources with constant surface charge densities with infinite spatial extent\footnote{ 
In the CGC approach this is called the McLerran-Venugopalan (MV) model.}.   
In this paper we restrict ourselves to the consideration of several different initializations involving asymmetric combinations of two transverse plane waves.

The initial condition that we use for the background field is
\bea
&& \varphi(\tau_0,\vec x_\perp) = \varphi_0\,\big(\cos(k_x x + k_y y )+ A \cos(k_y y)\big)
\, 
\label{init-c}
\eea
and the time derivative of the background field at $\tau=\tau_0$ is set to zero (numerically $10^{-3}$). 
The momenta $k_x$ and $k_y$ are chosen to take a range of values below the largest lattice eigenvalue, which is $\sqrt{8}$. 
The reader will note that our initial classical field is not always periodic and therefore does not respect our boundary conditions. The reason is that we want to avoid problems that may arise when resonant modes are considered, which in the present model would  correspond to the normal modes of the finite spatial lattice. For a large enough lattice all modes are effectively periodic and it is therefore expected that the precise form of the initial boundary conditions is not important.

\section{Physical observables}
\label{observ-sec}

\subsection{Canonical and conformal EMTs in Minkowski and Milne coordinates}
\label{compare-sec}

The EMT is  conserved
\bea
\nabla_\mu T^{\mu\nu} &=&   \partial_\mu T^{\mu\nu} + \Gamma^\mu_{\mu\sigma}T^{\sigma\nu}
+ \Gamma^\nu_{\mu\sigma}T^{\mu\sigma}=0\,
\label{div-cov}
\eea
which means that there is a  momentum four vector
 \bea
\PP^\nu=\int d\sigma_\mu T^{\mu\nu},
\eea
where $d\sigma_\mu$ is an element of a spacelike surface $\sigma$, whose components are conserved by the time evolution providing that the metric contains no explicit dependence on the coordinates $x^\mu$ and the fields vanish on the boundaries of $\sigma$.
This can be seen by noting that in  Minkowski coordinates the integral of the $\nu=0$ component of the conservation law (\ref{div-cov}) takes the form:
\bea
\int d^3 x \frac{\partial}{\partial t}T^{00} = -\int d^3 x\left(\partial_x T^{x0}+\partial_y T^{y0}+\partial_z T^{z0}\right)\,.
\eea
The right side is the integral of a total derivative so assuming the fields vanish at spatial infinity one finds that the energy defined as $E = \int d^3 x T^{00}$ is conserved ($dE/dt=0$). 

In Milne coordinates the situation is more complicated due to the $\tau$ dependence of the metric so that $\int d^3x \sqrt{-g}T^{00}$ integrated along a surface of constant $\tau$ is not in general conserved, even for an isolated system. 
In Milne coordinates the only non-zero components of the connection are $\Gamma^0_{\eta\eta} = \tau$ and $\Gamma^\eta_{0\eta} = \Gamma^\eta_{\eta 0} = 1/\tau$.
The $\nu=0$ component of the divergence equation  (\ref{div-cov}) gives
\bea
&&\frac{\partial T^{00}}{\partial\tau} =-\partial_\eta T^{\eta 0} - \frac{1}{\tau} T^{00}- \tau T^{\eta\eta} -\partial_x T^{x0}  -\partial_y T^{y0} \,.\label{div0} 
\eea
We  define the energy density and longitudinal and transverse pressures in terms of the  EMT in Milne coordinates as:
\bea
{\cal E} &=& \langle T^{00}\rangle \nn \\
 p_L&=&\tau^2 \langle T^{\eta\eta} \rangle \nn \\
p_T&=&\frac{1}{2}\langle T^{xx}+T^{yy}\rangle\,
\label{ep-defs}
\eea
where the angle brackets indicate an average over rapidity and the transverse coordinates. 
Since the average of a total derivative vanishes eq. (\ref{div0}) gives
\bea
&&\frac{\partial {\cal E}}{\partial \tau} = -\frac{1}{\tau}({\cal E} +p_L)
\,. \label{der1} 
\eea
We note that when the EMT is written in Milne coordinates $T^{00}$ does not correspond to the physical energy density as measured in the lab frame except at zero rapidity. 
In our calculation we include only a narrow slice of rapidity centered around mid-rapidity ($\eta=0$). The motivation is that we are trying to model the gluon fields produced in a relativistic collision of heavy ions for which the system of fields is largely rapidity independent at mid-rapidity. 
The definitions in (\ref{ep-defs}) also involve an average over this thin slice of rapidity. 
We therefore take the energy density and pressures as defined in (\ref{ep-defs}) as the physically relevant ones.

Now we address the fact that the canonical and conformal EMTs are not the same (even when they are calculated in the same coordinate system).
Equation (\ref{Tmunu-extra}) shows that the difference between the two EMTs, for all components, is a sum of terms that are either a total derivative in transverse coordinates, a total derivative in rapidity, or contain the product of the field and its time derivative (below we call the product $\phi \dot\phi$ a decoherence factor). As explained previously, the total derivative terms do not affect the bulk properties of the medium because they give zero when they are averaged.
The important point to understand when comparing the two EMTs is that the decoherence factor goes rapidly to zero at very early times. At $\tau=\tau_0$ we start with a fairly narrow Gaussian distribution of initial conditions centered on the chosen initial values $\varphi(\tau_0,\vec x_\perp)$ and $\dot\varphi(\tau_0,\vec x_\perp)$.  Each initial condition evolves independently using the equation of motion so that after some characteristic decoherence time the ensemble average of the product $\phi \dot\phi$ goes to zero. 
This means that the integrated EMT in (\ref{Tmunu-can}) is the same as the conformal form (\ref{Tmunu-cf}) except at very early times. If we only wanted to study late time properties of the system, like transverse-longitudinal pressure isotropization, both EMTs would be equally good. 
Since we are particularly interested in early time behaviour however, it is important to know which formalism is better for our purposes. 

To understand this we consider the trace of the two EMTs. 
The trace of the conformal EMT is zero by construction. The trace of the canonical EMT is
\bea
T^\mu_{~\mu}
&=&-\nabla^\mu\left(\phi\partial_\mu \phi \right) 
=   - \frac{d}{d\tau}(\varphi \dot\varphi) - \frac{\varphi \dot\varphi}{\tau}
+\vec\nabla_\perp \cdot(\varphi \vec\nabla_\perp\varphi)\,. 
\label{trace-can-milne}
\eea
This expression has terms that are proportional to the decoherence factor and its derivative, in addition to total derivatives, which means that the trace of the averaged canonical EMT is  not zero at  early times. 
We are interested in assessing the extent to which our system of fields behaves hydrodynamically, and the traditional formulation of hydrodynamics based on the assumption of local  equilibrium assumes a traceless EMT (see section \ref{hydro-sec}). 
This suggests we want to work with the conformal definition of the EMT. 
Furthermore, in section \ref{results-sec} we present some results that show that at early times the time dependence of the canonical EMT is characterized by rapid oscillations created by decoherence terms. These oscillations disappear at larger times but make it impossible to study early time dynamics. 
They are not present when the conformal EMT is used.

Finally we make the formal argument that we should work with a conformally invariant theory since we hope to use the scalar system to understand some properties of glasma.  

\subsection{Spatial and momentum asymmetries}

A correlation between initial spatial transverse asymmetry and azimuthal asymmetry in the momentum field is generally considered characteristic of the onset of hydrodynamic behaviour. Physically the idea is that if pressure gradients are converted into fluid velocities, then momentum anisotropy will grow as the spatial anisotropy decreases. 
Spatial deviations from azimuthal symmetry can be characterized with the quantity \cite{Kolb:1999it,Blaizot:2014nia}
\bea
&& \varepsilon = - 
\frac{\int d^2 x_\perp \left(\frac{x^2-y^2}{\sqrt{x^2 + y^2}}\right)T^{00}}{\int d^2 x_\perp {\sqrt{x^2 + y^2}}T^{00}} 
\label{ecc-def}
\eea
where we have used $\vec x_\perp = (x,y)$.
In a system of fields, momentum anisotropy can be described by an elliptic flow coefficient \cite{Carrington:2021qvi}
\bea
v_2 = \frac{\int d^2 x_\perp \left(\frac{(p^x)^2-(p^y)^2}{\sqrt{(p^x)^2 + (p^y)^2}}\right)}{\int d^2 x_\perp {\sqrt{(p^x)^2 + (p^y)^2}}}
\label{v2-def}
\eea
where $p^x = T^{0x}$ and $p^y = T^{0y}$. 
We comment that this definition is not directly related to what is normally measured in heavy-ion collisions, where one looks at the momentum distribution of produced particles. 
A different quantity that is commonly calculated in heavy-ion physics is the transverse pressure anisotropy defined 
\bea
A_{xy} = \frac{\int d^2 \vec x_\perp \,(T^{xx} - T^{yy})}{\int d^2 \vec x_\perp \,(T^{yy} + T^{xx})}\,.
\label{p-anio}
\eea
Hydrodynamic calculations show that $A_{xy}$ is closely related to the elliptic flow coefficient of produced particles \cite{Kolb:1999it} and the transverse pressure anisotropy has been used extensively in the literature to characterize momentum anisotropy \cite{Luzum:2008cw,Luzum:2009sb}.

\subsection{Universal flow}
\label{hydro-sec}

The authors of ref. \cite{Vredevoogd:2008id} derived an equation that describes the way that transverse flow develops for any system for which the longitudinal flow is boost invariant, the EMT is traceless, and the anisotropy of the spatial components of the EMT depends more strongly on $\tau$ than on the transverse coordinates at early times. In our calculation the first condition is satisfied by construction, the second is always true if we use the conformal EMT, and the third is well satisfied in our calculation.

We define the $x$-component of a velocity and an effective acceleration
\bea
&& V_x = \frac{T^{0x}}{T^{00}} \nn \\
&& \alpha = \frac{\partial V_x}{\partial\tau}  = \frac{\partial}{\partial \tau} \left( \frac{T^{0x}}{T^{00}} \right) = \frac{\partial_\tau T^{0x}}{T^{00}}-\frac{T^{0x}}{T^{00}}\frac{\partial_\tau T^{00}}{T^{00}}  \,.\label{alpha-def}
\eea
We can rewrite this in a more convenient form as follows. 
The zeroth component of the equation $\nabla_\mu T^{\mu\nu}$ in Milne coordinates is given in equation (\ref{div0}). The $x$-component is
\bea
&& \frac{\partial T^{0x}}{\partial\tau} = - \frac{1}{\tau}T^{0x} - \partial_x T^{xx} -\partial_y T^{yx} \,\label{divx} 
\eea
where we have dropped the derivative with respect to $\eta$ since we are interested in a system that it at least approximately boost invariant. We will assume that $T^{yx}$ is small and set it to zero.
We rewrite $\alpha$ using equations  (\ref{div0}, \ref{divx}) to obtain 
\bea
T^{00} \alpha &=& \partial_\tau T^{0x}- \frac{T^{0x}}{T^{00}} \partial_\tau T^{00} \nn \\
 &=& - \left(\frac{1}{\tau}T^{0x} + \partial_x T^{xx} \right)
+ \frac{T^{0x}}{T^{00}} \left(\frac{1}{\tau} T^{00} + \tau T^{\eta\eta} + \partial_x T^{x0}  + \partial_y T^{y0}\right)\,. \label{temp1}
\eea

We assume that at early times the field $\phi$ has a power series expansion in $\tau$ and define
\bea
\phi  = \sum_{i=0}\phi_{2i} \tau^{2i}
\eea
where only even powers of $\tau$ are included because the equation of motion does not allow solutions with odd powers. From the explicit expressions for the components of the EMT (\ref{Tmunu-comps}, \ref{Tmunu-extra}) we see that the diagonal components are series with only even powers of $\tau$ and both $T^{x0}$ and $T^{y0}$ have only odd powers. We define $T^{00} = T^{00}_0 + T^{00}_2 \tau^2 + \dots$ where the subscript indicates the power of $\tau$ associated with each coefficient, and similarly for each component of the EMT. 

We write $\alpha = \alpha_0  + \dots$ where $\alpha_0 = T^{0x}_1/T^{00}_0$ and keep terms in (\ref{temp1}) to order $\tau^0$ which gives
\bea
T^{0x}_1 = - \partial_x T^{xx}_0 - T_1^{0x} + \frac{T_1^{0x}}{T_0^{00}}(T^{00}_0 + T^{\eta\eta}_{-2} )\,.
\eea
Rearranging we have
\bea
T_1^{0x}(T^{00}_0 - T^{\eta\eta}_{-2}) =-T^{00}_0 \partial_x T^{xx}_0\,. 
\label{tempb}
\eea
The condition that the trace is zero, to zeroth order in $\tau$, is $T^{00}_0-T^{\eta\eta}_{-2} -T^{xx}-T^{yy}=0$. Equation (\ref{tempb}) therefore takes the form 
\bea
T_1^{0x} = -\frac{T^{00}_0}{T^{xx}_0+T^{yy}_0} \partial_x T^{xx}_0\,.
\label{tempc}
\eea
We assume that at early times $T^{xx} \approx T^{yy}$ and that $T^{00}/T^{xx}$ is approximately independent of $\vec x_\perp$. Using these assumptions and multiplying by $\tau$ equation (\ref{tempc}) gives
\bea
T^{0x} = -\frac{\tau}{2}  \partial_x T^{00} + {\cal O}(\tau^2) \,
\label{hydro-1b}
\eea
or equivalently
\bea
\frac{\partial v_x}{\partial \tau} = -\frac{1}{2}  \frac{\partial_x T^{00}}{T^{00}} + {\cal O}(\tau^2) \,.
\label{hydro-1c}
\eea
This result tells us that at early times the impulse to the transverse collective flow is connected to the gradient of the energy density. It means that different systems with the same energy density but very different pressures, ranging from hydrodynamic $p_T = p_L = {\cal E}/3$ to systems with highly asymmetric initial pressures like ours, will develop transverse flow in the same way. 
In section (\ref{results-sec}) we show that (\ref{hydro-1b}) is well satisfied in our calculation, at early times. 

For comparison we discuss how (\ref{hydro-1b}) can be derived for a system governed by hydrodynamics. 
The hydrodynamic definition of the EMT in Minkowski coordinates is
\bea
T^{\mu\nu}_{\rm hydro} = (\epsilon +p)u^\mu u^\nu -p g^{\mu\nu} \,
\label{T-hydro}
\eea
where $\epsilon$ is the energy density, $p$ is the thermodynamic pressure, and $u^\mu = \gamma(1,\vec v)$ is the fluid velocity. The tracelessness of the EMT gives the equation of state $p=\epsilon/3$. 
If we assume that each velocity component can be written as a power series in $t$
\bea
v^i=v^i_0+v^i_1 t+v^i_2 t^2 +\dots
\eea
then at leading order we find
\bea
T^{0x} = -t\frac{\partial T^{xx}}{\partial x} +{\cal O}(t^2)\,
\label{hydro-mink}
\eea
and since $T^{xx}=p=\epsilon/3$ we have  
\bea
T^{0x} = -\frac{t}{3}\frac{\partial T^{00}}{\partial x} +{\cal O}(t^2)\,.
\label{hydro-mink-3d}
\eea
The same calculation in two spatial dimensions gives
\bea
T^{x0} = -\frac{t}{2}\frac{\partial T^{00}}{\partial x} +{\cal O}(t^2)\,
\label{hydro-mink-2d}
\eea
which agrees with (\ref{hydro-1b}) at mid-rapidity.

\section{Results}
\label{results-sec}

Our goal is to study how the system isotropizes. We are particularly interested in understanding the extent to which the system behaves hydrodynamically, and how hydrodynamic behaviour develops with time. Unless stated otherwise, all calculations are done with $(N,L,N_\gamma)=(121,41,256)$. We have checked that results are not sensitive to these numbers. The initialization is specified by giving the values of $(\varphi_0,A,k_x,k_y)$ that are used in equation (\ref{init-c}). The coupling is $g=4$ in all calculations. 

\subsection{Trace of the canonical EMT}

\begin{figure}[H]
\centering
\includegraphics[scale=0.95]{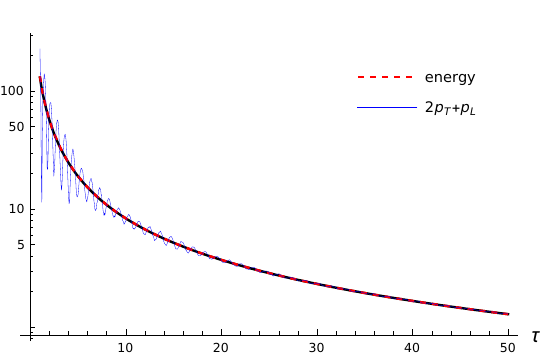}
\caption{The energy (red) and the sum $p_L+2p_T$ (blue) for the canonical EMT. For the conformal EMT the sum of the pressures and the energy are right on top of the canonical energy curve.  The calculation is done with $(\varphi_0,A)=(15,0)$ and $k_x=k_y=0.41$. 
\label{plot-trace}}
\end{figure}

The trace of the averaged canonical EMT is not zero, as shown in equation (\ref{trace-can}), but it goes rapidly to very small values as the system evolves because the decoherence factor goes to zero. 
In fig. \ref{plot-trace} we show the energy and the sum of the pressures for the  canonical EMT.  The blue curve, which shows the sum of the pressures, oscillates around the red curve, which shows the energy, until  $\tau \approx 20$. At larger times the trace (the difference between the blue and red lines) is numerically very close to zero. 
For the conformal EMT the sum of the pressures and the energy are right on top of the canonical energy curve.  
The oscillations in the pressures from the canonical EMT are typical of the early time behaviour and caused by the decoherence terms. 

\subsection{Transverse-Longitudinal Isotropization}

A simple way to study how the system comes to equilibrium is to see how closely it satisfies the isotropic equation of state: $p_L=p_T = {\cal E}/3$. 
This was studied by DEGV who showed that $(p_T-p_L)/{\cal E}$ decreases as a function of time\footnote{
Note that because we work numerically with a box of finite size in coordinate space, the calculation will break down at large times because there will be highly occupied momentum modes that are not supported.}.  
The result is important because it proves that the resummation method developed by DEGV is successful in capturing the dominant physics of the expanding plasma. 

\begin{figure}[H]
\centering
\includegraphics[scale=1.25]{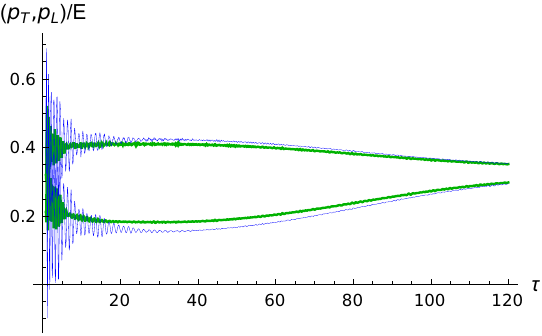}
\caption{The transverse and longitudinal pressures normalized by the energy from the canonical EMT  for the initializations $(\varphi_0,A)=(15,0)$ and $(\varphi_0,A)=(15/\sqrt{2},1)$. Both calculations are done with $k_x=k_y=0.41$. In both cases the upper curves are $p_T/{\cal E}$ and the lower lines are $p_L/{\cal E}$.
\label{plot-rt2}}
\end{figure}

Figure \ref{plot-rt2} shows the transverse and longitudinal pressures normalized by the energy from the canonical EMT, for two initializations that have the same initial field amplitude. The figure shows that the isotropization of the transverse and longitudinal pressures is not significantly affected when the second cosine is included in the initialization. 

\begin{figure}[H]
\centering
\includegraphics[scale=1.25]{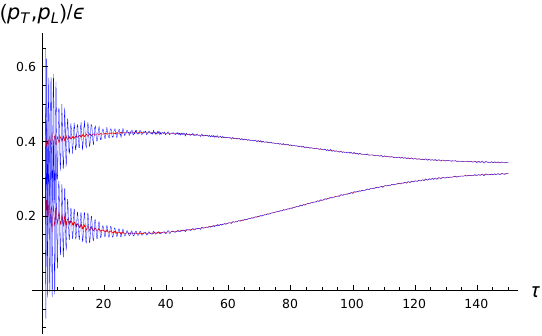}
\caption{The normalized transverse and longitudinal pressures for the canonical EMT (blue) and the conformal EMT (red).  The initialization is $(\varphi_0,A) = (15,0)$ and $k_x=k_y=0.41$. The upper and lower lines are respectively $p_T/{\cal E}$ and $p_L/{\cal E}$.
\label{plot-pTpL}}
\end{figure}

In fig. \ref{plot-pTpL} we show the transverse and longitudinal pressures normalized by the energy for the canonical and conformal EMTs. The blue curves show the oscillations that are characteristic of the canonical EMT, which die off at larger times when the decoherence terms $\sim \phi\dot\phi$ go to zero. The red curves are the results from the conformal EMT and are much smoother at early times. At $\tau \gtrsim 40$ the results from the two EMTs are indistinguishable. 

\subsection{Azimuthal asymmetry}

\begin{figure}[H]
\centering
\includegraphics[scale=0.88]{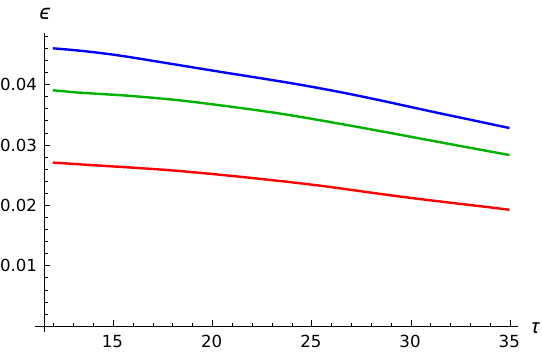}
\includegraphics[scale=0.88]{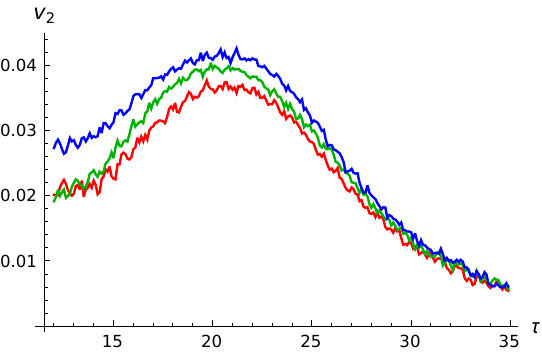}
\caption{The left panel shows the eccentricity as a function of $\tau$ and the right panel is the elliptic flow coefficient $v_2$. The initialization is $(\varphi_0,A,k_x)=(15,1,1.47)$ and from bottom to top $k_y=(1.12\text{~(red)},1.30 \text{~(green)},1.47\text{~(blue)})$.  The calculations are done with $(N,L,N_\gamma)=(81,31,256)$. 
\label{plot-EV}}
\end{figure}

\begin{figure}[H]
\centering
\includegraphics[scale=0.88]{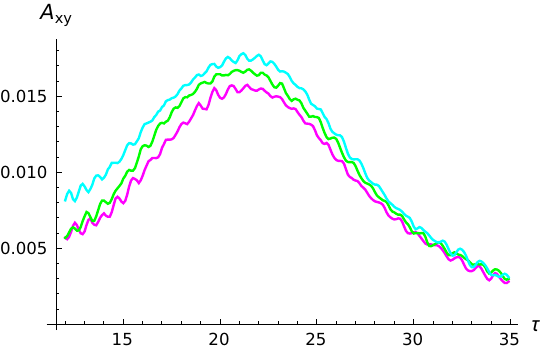}
\includegraphics[scale=0.88]{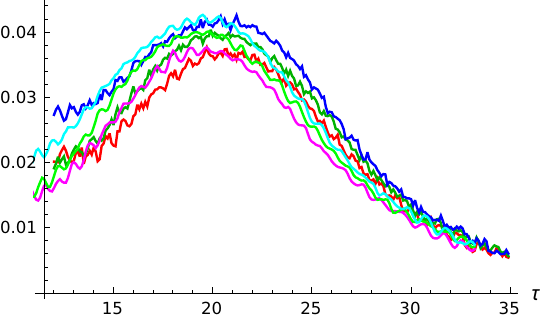}
\caption{The left panel is the transverse pressure asymmetry in equation (\ref{p-anio}) for the same initializations as in fig. \ref{plot-EV}. The right panel is $v_2$ superimposed on a plot of $A_{xy}$ that has been shifted left 1.7 time units and multiplied by a factor of 2.4. The lighter coloured lines show the shifted results for $A_{xy}$. 
\label{plot-Axy}}
\end{figure}

When we use an initialization with $A\ne 0$ the initial fields are azimuthally asymmetric. 
We want to study how the azimuthal asymmetry of the energy density and momentum distribution develop in time, and how they are related to each other. 
The asymmetry in the transverse plane is orders of magnitude smaller than the transverse-longitudinal asymmetry discussed in the previous subsection, which makes it hard to study numerically. 
In fig. \ref{plot-EV} we show the eccentricity (left panel) and elliptic flow coefficient  (right panel) as functions of $\tau$ for initializations $(\varphi_0,A,k_x)=(15,1,2.83)$ and $k_y\in(1.30,1.47,2.83)$. 
One sees that the initial spatial eccentricity decreases with time, and that the size of the initial eccentricity is correlated with the size of the momentum anisotropy that is produced. The elliptic flow coefficient initially increases and eventually decays. 
In  fig. \ref{plot-Axy} we show the similarity in the shape of the two measures $v_2$ and $A_{xy}$ defined in equations (\ref{v2-def}, \ref{p-anio}). The left panel shows $A_{xy}$ for the same three initializations as in fig. \ref{plot-EV}. The right panel shows the elliptic flow coefficient, and the transverse pressure asymmetry shifted left by 1.7 time units and multiplied by 2.4. 

\subsection{Universal flow}

\begin{figure}[H]
\centering
\includegraphics[scale=0.85]{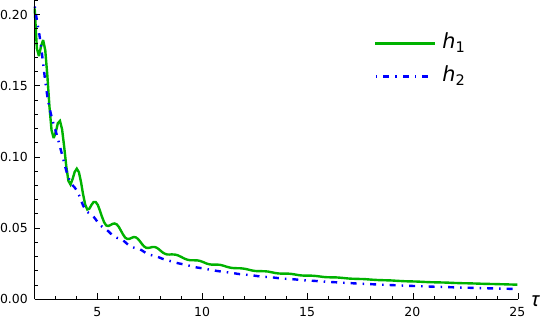}
\includegraphics[scale=0.85]{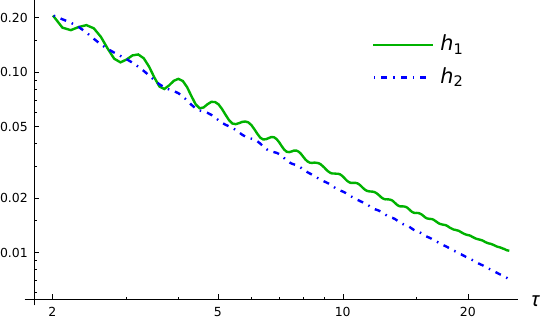}
\caption{The two quantities in equation (\ref{hydro-compare}) at early times. The left panel is a linear plot and the right is a log-log graph that shows the differences at later times more clearly. The calculation is done with $(\varphi_0,A)=(15,0)$ and $k_x=k_y=0.41$. 
\label{plot-hydro}}
\end{figure}

We can check how well equation (\ref{hydro-1b}) is satisfied for our system of scalar fields. 
The calculation is numerically difficult because the functions on both sides of the equation fluctuate in transverse position at early times, and decay rapidly as time increases. To compare them we  calculate the root of the average of the squares of both sides of the equation. We define
\bea
&&h_1 = \sqrt{ \left\langle \left(\frac{\tau}{2} \frac{\partial T^{00}}{\partial x}\right)^2\right\rangle} \text{~~and~~}
h_2 =  \sqrt{\langle (T^{0x})^2\rangle} 
\label{hydro-compare}
\eea
where the angle brackets indicate an average over spatial rapidity and the transverse coordinates\footnote{The factor $\tau$ in $h_1$ is multiplied by the length scale $1/L$ because we have set the lattice spacing in the transverse direction to one.}.  Figure \ref{plot-hydro} shows there is good agreement of the two quantities in (\ref{hydro-compare}) at early times. 

\section{Conclusions}
\label{sec-conclusions}

In this paper we have studied the dynamics of a self-interacting real scalar field. 
We work with a conformally invariant action and calculate the energy momentum tensor using the classical statistical approximation. 
We have studied the time evolution of the azimuthal asymmetry  of the energy density and the momentum field and shown that the two quantities are correlated. The eccentricity decreases and the momentum elliptic flow coefficient initially increases and then later decays. 
We have studied the asymmetry of the transverse pressures and shown that its behaviour is strikingly similar to that of the elliptic flow coefficient. 
Both of these behaviours are seen in hydrodynamic calculations that describe the physics of a relativistic heavy-ion collision at much later times.  
We have compared the derivative of the transverse velocity and the gradient of the energy density and shown that the characteristics of universal flow are present in our system. 

Our results provide further support for the use of the classical statistical approximation to describe a system of strongly coupled fields and
suggest that the method is worthy of further investigation using more realistic initial configurations, and physical theories that involve additional fields. 
Such calculations could lead to a deeper understanding of the early onset of hydrodynamic behaviour that is seen in heavy ion collisions.

\appendix

\section{Useful equations}
\label{A-sec}

In Milne coordinates the only non-zero components of the connection are $\Gamma^0_{\eta\eta} = \tau$ and $\Gamma^\eta_{0\eta} = \Gamma^\eta_{\eta 0} = 1/\tau$.
The transformation from Minkowski to Milne coordinates is 
 \bea
A^{\mu}{}_{\nu} \equiv \frac{\partial x^{\mu}}{\partial x^\nu} &=&
 \left[
\begin{tabular}{c c c c}
$\cosh{\eta}$ & $-\sinh(\eta)$ & 0 & 0\\
$-\frac{1}{\tau}\sinh(\eta)$& $\frac{1}{\tau}\cosh(\eta)$ & 0 & 0 \\
0 & 0 & -1 & 0 \\
0 & 0  & 0 & -1 \\
\end{tabular}  
\right]\,.
 \eea
Some results for the gradient operator are
\bea
&& \partial_\mu  = (\partial_\tau,\partial_\eta ,\vec\nabla) \nn \\
&& \partial^\mu  = (\partial_\tau,-\frac{1}{\tau^2}\partial_\eta ,-\vec\nabla) \nn \\
&& (\partial_\mu \varphi) (\partial^\mu \varphi) =\dot\varphi^2-\frac{1}{\tau^2}(\partial_\eta\varphi)^2 - (\vec\nabla\varphi)^2 \nn \\
&& \Box  = \partial_\tau^2+\frac{1}{\tau}\partial_\tau - \frac{1}{\tau^2}\partial_\eta^2 -\vec\nabla^2 \,.
\eea

In equations (\ref{Tmunu-comps}, \ref{Tmunu-extra}) we give the components of the canonical EMT (\ref{Tmunu-can}) and the additional piece that appears in the conformal EMT (\ref{Tmunu-ex}).
For the diagonal terms we give two ways to write the components of $T^{\mu\nu}_{\rm ex}$. 
The label `alt' indicates that the equation of motion has been used. These expressions are easier to calculate numerically. The components of the canonical EMT are
\bea
&& T_{\rm can}^{00} = \frac{1}{2}\left((\dot \phi)^2 +\frac{(\partial_\eta\phi)^2}{\tau^2} +(\partial_x\phi)^2+(\partial_y\phi)^2\right)+\frac{g^2}{4!}\phi^4\nn\\
&& T_{\rm can}^{\eta\eta} =\frac{1}{2}\tau^{-4}(\partial_\eta\phi)^2+\frac{1}{\tau^{2}}\left[\frac{1}{2}\left((\dot \phi)^2 - (\partial_x\phi)^2-(\partial_y\phi)^2\right)-\frac{g^2}{4!}\phi^4\right]\nn\\
&& T_{\rm can}^{xx} = \left[\frac{1}{2}\left((\dot \phi)^2 -\frac{(\partial_\eta\phi)^2}{\tau^2} +(\partial_x\phi)^2-(\partial_y\phi)^2\right)-\frac{g^2}{4!}\phi^4\right]\nn\\
&& T_{\rm can}^{yy}=\left[\frac{1}{2}\left((\dot \phi)^2 -\frac{(\partial_\eta\phi)^2}{\tau^2}-(\partial_x\phi)^2+(\partial_y\phi)^2\right)-\frac{\lambda}{4!}\phi^4\right]\nn\\
&& T_{\rm can}^{xy} = (\partial_x\phi)(\partial_y\phi)\nn \\
&& T_{\rm can}^{0\eta}= - \frac{1}{\tau^2} \dot\phi(\partial_\eta\phi)  \nn\\
&& T_{\rm can}^{0y} = -\dot\phi(\partial_y\phi)\nn \\
&& T_{\rm can}^{0x} =  -\dot\phi(\partial_x\phi) \,. \label{Tmunu-comps}
\eea
The components of $T^{\mu\nu}_{\rm ex}$ are 
\bea
&& T_{\rm ex}^{00} = \frac{1}{3}\frac{1}{\tau} (\varphi\dot\varphi) 
-\frac{1}{3} \frac{1}{\tau^2}\big((\partial_\eta \varphi)^2 + \varphi \partial_\eta^2 \varphi \big)
-\frac{1}{3} \vec\nabla_\perp \cdot(\varphi \vec \nabla_\perp \varphi) \nn\\
&& T_{\rm ex-alt}^{00} = -\frac{1}{3}\left[\phi \ddot\phi + \frac{g^2}{6}\phi^4 +\frac{(\partial_\eta\phi)^2}{\tau^2}
+|\nabla_\perp \phi|^2 \right]  \nn\\
&& T^{\eta\eta}_{\rm ex} = -\frac{1}{3\tau^2}\frac{\partial}{\partial \tau} (\varphi\dot\varphi) +\frac{1}{3\tau^2} \vec\nabla_\perp \cdot(\varphi \vec \nabla_\perp \varphi)\nn \\
&& T_{\rm ex-alt}^{\eta\eta} =-\frac{1}{3\tau^4}\left[\phi\partial_\eta^2 \phi - \tau\phi\dot\phi -\frac{g^2}{6}\phi^4\tau^2 
+\tau^2\left(\dot\phi^2-|\nabla_\perp\phi|^2\right)\right]  \nn \\
&& T^{xx}_{\rm ex} = -\frac{1}{3}\left(\frac{\partial}{\partial \tau} (\varphi\dot\varphi) + \frac{1}{\tau} (\varphi\dot\varphi) -\frac{1}{\tau^2} \partial_\eta(\varphi\partial_\eta\varphi) \right) 
+\frac{1}{3} \frac{\partial}{\partial y}\left(\phi \frac{\partial \phi}{\partial y}\right)\,\nn \\
&& T_{\rm ex-alt}^{xx} = -\frac{1}{3}\left[\phi\partial_x^2\phi -\frac{g^2}{6}\phi^4 +(\dot\phi)^2 -\frac{(\partial_\eta\phi)^2}{\tau^2} -(\partial_y\phi)^2\right] \nn \\
&& T^{yy}_{\rm ex} = -\frac{1}{3}\left(\frac{\partial}{\partial \tau} (\varphi\dot\varphi) + \frac{1}{\tau} (\varphi\dot\varphi) -\frac{1}{\tau^2} \partial_\eta(\varphi\partial_\eta\varphi) \right) 
+\frac{1}{3} \frac{\partial}{\partial x}\left(\phi \frac{\partial \phi}{\partial x}\right)\,  \nn \\
&& T_{\rm ex-alt}^{yy} = -\frac{1}{3}\left[\phi\partial_y^2\phi -\frac{g^2}{6}\phi^4 +(\dot\phi)^2 -\frac{(\partial_\eta\phi)^2}{\tau^2} -(\partial_x\phi)^2\right] \nn \\
&& T_{\rm ex}^{0\eta} = \frac{1}{3\tau^2}\left(\phi\partial_\eta\dot\phi + \dot\phi\partial_\eta\phi - \frac{1}{\tau}\phi\partial_\eta\phi\right )  \nn \\
&& T_{\rm ex}^{0x} = \frac{1}{3}(\phi\partial_x\dot\phi+ \dot\phi \partial_x\phi)  \nn \\
&& T_{\rm ex}^{0y} = \frac{1}{3}(\phi\partial_y\dot\phi+ \dot\phi \partial_y\phi)  \nn \\
&& T_{\rm ex}^{xy} =-\frac{1}{3} \left[\phi(\partial_x\partial_y\phi) + (\partial_x\phi)(\partial_y\phi)\right] \,. \label{Tmunu-extra}
\eea

\begin{acknowledgments}
Margaret Carrington gratefully acknowledges helpful discussions with Fran\c{c}ois Gelis. 

This work has been supported by the Natural Sciences and
Engineering Research Council of Canada Discovery Grant program 
 from grants 2023-00023 and 2018-04090.
This research was enabled in part by support provided by WestGrid
(www.westgrid.ca) and the Digital Research Alliance of Canada (alliancecan.ca).

\end{acknowledgments}


\begin{thebibliography}{}

\bibitem{Dusling:2010rm}
K.~Dusling, T.~Epelbaum, F.~Gelis and R.~Venugopalan,
Nucl. Phys. A \textbf{850}, 69, (2011).
\bibitem{Epelbaum:2011pc}
T.~Epelbaum and F.~Gelis,
Nucl. Phys. A \textbf{872}, 210 (2011).
\bibitem{Dusling:2012ig}
K.~Dusling, T.~Epelbaum, F.~Gelis and R.~Venugopalan,
Phys. Rev. D \textbf{86}, 085040, (2012).

\bibitem{Carrington:2022ccm}
M.~E.~Carrington, G.~Kunstatter, C.~D.~Phillips and M.~E.~Rubio,
Entropy \textbf{24}, 1612 (2022). 

\bibitem{Berges:2013lsa}
J.~Berges, K.~Boguslavski, S.~Schlichting and R.~Venugopalan,
JHEP \textbf{05}, 054, (2014).
\bibitem{Epelbaum:2014yja}
T.~Epelbaum, F.~Gelis and B.~Wu,
Phys. Rev. D \textbf{90}, 065029, (2014). 
\bibitem{Gelis:2024iar}
F.~Gelis and S.~Hauksson,
[arXiv:2403.11908 [hep-ph]].

\bibitem{Vredevoogd:2008id}
J.~Vredevoogd and S.~Pratt,
Phys. Rev. C \textbf{79}, 044915, (2009).

\bibitem{Blaschke:2016ohs}
D.~N.~Blaschke, F.~Gieres, M.~Reboud and M.~Schweda,
Nucl. Phys. B \textbf{912}, 192, (2016).

\bibitem{PT} Leonard Parker and David Toms, Quantum Field Theory in Curved Spacetime, Cambridge University Press, 2009.

\bibitem{Kolb:1999it}
P.~F.~Kolb, J.~Sollfrank and U.~W.~Heinz,
Phys. Lett. B \textbf{459}, 667 (1999).

\bibitem{Blaizot:2014nia}
J.~P.~Blaizot, W.~Broniowski and J.~Y.~Ollitrault,
Phys. Lett. B \textbf{738}, 166 (2014).

\bibitem{Carrington:2021qvi}
M.~E.~Carrington, A.~Czajka and S.~Mr\'owczy\'nski,
Phys. Rev. C \textbf{106}, 034904, (2022). 

\bibitem{ollitrault-1}
J.Y. Ollitrault, Phys. Rev. D \textbf{46}, 229, (1992).

\bibitem{Luzum:2008cw}
M.~Luzum and P.~Romatschke,
Phys. Rev. C \textbf{78}, 034915, (2008).

\bibitem{Luzum:2009sb}
M.~Luzum and P.~Romatschke,
Phys. Rev. Lett. \textbf{103}, 262302, (2009).


\end{thebibliography}
\end{document}